\documentclass[prl,aps,twocolumn,superscriptaddress,showpacs]{revtex4}
\usepackage{amsmath}
\usepackage{amsbsy}
\usepackage{amssymb}
\usepackage[dvips]{graphicx}

\begin{document}
\title{A frozen glass phase in the multi-index matching problem}

\author{O. C. Martin}
\affiliation{Laboratoire de Physique Th\'eorique et Mod\`eles Statistiques,
b\^atiment 100, Universit\'e Paris-Sud, F--91405 Orsay, France.}

\author{M. M\'ezard}
\affiliation{Laboratoire de Physique Th\'eorique et Mod\`eles Statistiques,
b\^atiment 100, Universit\'e Paris-Sud, F--91405 Orsay, France.}

\author{O. Rivoire}
\affiliation{Laboratoire de Physique Th\'eorique et Mod\`eles Statistiques,
b\^atiment 100, Universit\'e Paris-Sud, F--91405 Orsay, France.}

\date{\today}

\begin{abstract}
The multi-index matching is an NP-hard combinatorial optimization
problem; for two indices it reduces to the well understood bipartite
matching problem that belongs to the polynomial complexity class. We use
the cavity method to solve the thermodynamics of the multi-index system
with random costs.
The phase diagram is much richer than for the case of the bipartite matching
problem: it shows a finite temperature phase transition to a completely
frozen glass phase, similar to what happens in the random energy model.
We derive the critical temperature, the ground state energy density, and
properties of the energy landscape, and compare the results to 
numerical studies based on exact analysis of small systems.
\end{abstract}

\maketitle

It has been recognized  early on that one important motivation of
the research in spin glass theory is the ubiquity of systems with frustration and
disorder~\cite{MezardParisi87b}. In particular, recent statistical physics
studies have brought interesting new results in some important computer
science problems. Notable examples are found in optimization (e.g.
$K$-satisfiability ($K$-SAT)~\cite{MezardParisi02}, graph
coloring~\cite{MuletPagnani02}, or vertex cover~\cite{WeigtHartmann03b}) and
information theory (e.g. error correcting codes~\cite{Montanari01}). So
far, the most interesting applications of spin glass theory have been obtained in
this emerging field, 
which witnesses an upsurge of interdisciplinary studies
involving physicists, computer scientists, and probabilists.

One of the first optimization problems studied analytically by physics methods
was the random Bipartite Matching Problem (BMP). It is also a simple
problem:  from the computer science point of view, it belongs to the class P
of polynomial complexity; from the physics point of view, it has no phase
transition at finite temperature, and its solution with the replica
method~\cite{MezardParisi85} shows a simple replica symmetric behavior.
Interestingly, the validity of this solution has been recently confirmed by a
rigorous mathematical study~\cite{Aldous01}.
 
In this work we study the Multi-Index Matching Problem (MIMP), a natural
extension~\cite{Pierskalla68} of the BMP. This is a more complicated problem:
it belongs to the NP-hard class, and as we will see it also exhibits a finite
temperature phase transition, with a low temperature glassy phase.
Using the cavity method, we find that this phase consists
of isolated configurations, and we conjecture that our method yields
an exact solution to this problem. Because of its structural resemblance to
the BMP, one may hope that the MIMP  will also be amenable to rigorous study,
generalizing the construction of~\cite{Aldous01} to a problem with a 
glass phase.

\paragraph*{The random MIMP ---} 
In  the BMP one is given two sets of $M$ points,  $S_1$ and  $S_2$.
Each point of  $S_1$ must be ``matched'' or assigned to one point
of $S_2$. This matching must be one-to-one, and
it can be represented by the ``occupation'' of the edges 
between the points of the two sets; we define $n_{i_i,i_2}=1$
if the points $(i_1,i_2)\in S_1\times S_2$ are matched, while
$n_{i_1,i_2}=0$ otherwise. To each matching we associate
a cost or energy, which is the sum of the costs
of each occupied edge.

The MIMP is a straightforward generalization of this  problem
to more than two indices. Given $d$ sets
$S_1$,\dots,$S_d$, each of $M$ sites, 
a hyperedge is a $d$-uplet where
exactly one site from each set appears. For
each hyperedge we introduce a cost $\ell_{i_1,\dots, i_d}$, 
and the total cost of a (multi-index) matching is given, 
in terms of the occupation numbers of hyperedges, by:
\begin{equation}
\label{eq:Hhypermatching}
H[\{n_{i_1,\dots, i_d}\}] =
\sum_{i_1,\dots, i_d} \ell_{i_1,\ldots,i_d}n_{i_1,\dots,i_d}   \ .
\end{equation}
The occupation numbers of hyperedges, $n_{i_1,\dots,i_d}\in \{0,1\} $ 
must be such that each site appears exactly once:
\begin{equation}\label{eq:constraints}
\forall r\in[1,d],\quad \forall i_r,\quad \sum_{i_1,\dots,i_{r-1},i_{r+1},\dots,i_d}n_{i_1,\dots,i_d}=1.
\end{equation}

The optimization problem consists in finding the minimum cost matching. What
makes this problem difficult is the constraint~(\ref{eq:constraints}) of
having each site appear exactly in just one hyperedge; for $d \ge 3$ the MIMP
is NP-hard~\cite{Karp72}. MIMP arise for instance when assigning tasks (jobs)
to people in particular time slots or in different places. An application also
arises in the context of track reconstruction~\cite{Poore94}: given the
positions of $M$ unlabeled particles at $d$ different times, one is to
determine the tracks or trajectories of each. This kind of formulation is in
fact used in track reconstruction in high energy
physics~\cite{PusztaszeriRensing95}.

We shall study the random MIMP where the individual costs $\ell_{i_1,\dots
i_d}$ are independent identically distributed random variables. For
definiteness we shall take $\ell_{i_1,\dots i_d}$ to have uniform distribution
in $\left[0,1\right]$, although other distributions can be studied similarly.

 For a given
sample $\ell$, characterized by the values of $\ell_{i_1,\dots i_d}$,
 the
partition function at inverse temperature ${\tilde\beta}\equiv T^{-1}$ is
\begin{equation}
\label{eq:Zhypermatching}
Z_{\ell}=\sum_{\{n_{i_1,\dots,i_d}\}}
e^{-{\tilde\beta} H[\{n_{i_1,\dots,i_d}\}] }
\end{equation}
where $\tilde{\beta}$ is the inverse temperature and the sum is 
over all possible matchings.
In the thermodynamic limit
($M \to \infty$), only the behavior at the 
lowest values of $\ell_{i_1,\dots,i_d}$ matters. Indeed, if we consider a given
site in any of the $d$ sets, it is to be assigned to a low 
cost hyperedge; generally it is possible to
assign it to one of the first shortest such hyperedges. This
means that at large $M$, the typical cost of an occupied hyperedge
in the low temperature regime should scale as $1/M^{d-1}$.
It is thus convenient to work with rescaled quantities
that are extensive (i.e. proportional to $M$): 
\begin{equation}
\label{eq:Ehypermatching}
E = M^{d-1} H ~ .
\end{equation}
This amounts to considering thermodynamic quantities 
and having $\beta = \tilde{\beta} / M^{d-1}$ as the control parameter:
one should keep $\beta$ fixed when taking the large $M$ limit.

Given these considerations, we conjecture that the free energy
density is
self-averaging as in most disordered systems, and in particular as rigorously
proved for $d=2$ \cite{Aldous90}.

\paragraph*{Numerical study of the ground state ---}
For a given sample of the quenched disorder, we want to determine the ground
 state energy $E_0$ which is the minimum of all $E[\{n_{i_1,\dots i_d} \}]$.
 An exhaustive search over all matchings works only for very small $M$
 (typically $M\le 6$ when $d \ge 3$) because of the rapid growth of the number
 of legal matchings, in $(M!)^{d-1}$. We have followed instead a branch and
 bound (B\&B) approach which allows us to study intermediate $M$. From such an
 algorithm, we can test numerically whether $E_0$ is self-averaging and study
 its large $M$ limit.

The determination of the best matching uses a search tree. All the nodes at
level $p$ of this tree correspond to having chosen hyperedges for the first
$p$ points of the set $S_1$ (ordered arbitrarily). To go from level $p$ to
level ($p+1$), we branch on all possible $M^{d-1}$ choices for the next
hyperedge. Then a path from the tree's root (level 0) to a leaf (level $M$) is a choice of $M$ hyperedges which may or not correspond to a legal matching.
The cost of a node in the tree is defined as: the sum of the costs of its
associated hyperedges when they don't overlap, or $\infty$ if the hyperedges
overlap (i.e. they use a point of the $d$ sets $S_i$ more than once).

The B\&B algorithm searches the tree and implements
pruning. For this, it needs an upper bound $E_{\rm ub}$ on $E_0$; 
we initialize this quantity before performing the search 
via the cost of a legal matching  obtained by a greedy 
assignment of the hyperedges. Then the algorithm starts at the
root of the tree (level 0) and searches it recursively. At each
level, one branches on the $M^{d-1}$ choices of hyperedges
that take one to the next level. Every time the current
node has a cost greater than $E_{\rm ub}$, all of its descendent nodes
can be ignored as they cannot contain the ground state. If we reach
level $M$, we have a legal matching which we keep if its cost
is less than $E_{\rm ub}$ (and we update $E_{\rm ub}$ accordingly).
Upon termination, we have the ground state and $E_0=E_{\rm ub}$.

\begin{figure}
\includegraphics[width=8cm, height=5cm,angle=0]{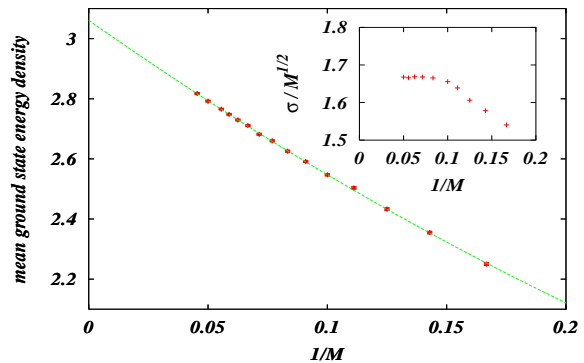}
\caption{Mean ground state energy density ${\overline E_0}/M$ 
as a function of $1/M$ 
in the 3-index problem and our best fit. Inset: standard deviation $\sigma(E_0)/\sqrt{M}$
versus $1/M$.
\protect\label{fig:e0}} 
\end{figure}

We have implemented this algorithm along with a number of optimizations.
For our computer program, one $d=3$ sample at $M=20$ takes typically 
5 seconds on a 2 GHz PC, and the CPU time
grows by a factor around 2.2 every time $M$ is increased by 1.
We have performed runs 
for $M \le 22$ with 20000  samples at each $M$. From 
these data, we have extracted $\overline{E_0}$, the disorder average 
of $E_0$; the mean cost per node 
is shown in Fig.~\ref{fig:e0}. The data for $M\ge 10$
are well fitted by a
quadratic curve in $1/M$, giving
$\overline{E_0}/M \to 3.06 \pm 0.03$;  a power law fit of the same
quality gives  $\overline{E_0}/M \to 3.09 \pm 0.03$.

In the inset of the figure, we show that the standard
deviation $\sigma(E_0/M)$ scales as $1/\sqrt{M}$, which
is evidence for self-averaging and also suggests a central limit
theorem behavior when $M \to \infty$.

Finally, we have also investigated a bit the case of $d=4$; however,
we were limited to $M \le 15$ and used only 5000 samples.
(The CPU time grows by about the same factor
when $M$ is increased by 1 as when $d=3$.)
Our best fit in this case leads to $\overline{E_0}/M \to 7.2(3)$.

\paragraph*{Thermodynamics and the cavity approach ---}
The recent formulation of the cavity method~\cite{MezardParisi01} for diluted
systems offers a choice tool to study the thermodynamics of the MIMP
analytically. Building on the idea that the optimal matching selects
preferentially the hyperedges with the lowest costs, we dilute the initially
complete hypergraph by suppressing hyperedges with $\ell_{i_1,...,i_d}> C
M^{1-d}$ \cite{footnote1}. In the resulting graph, the degree of each site is a
Poisson distributed variable of mean $C$. When increasing $M$ to $M+1$, a new
serie of $d$ sites is added. Each of them is connected to a finite number of
neighbours. The partition function of one new site is easily computed in terms
of the probability, $\exp[\beta (x_i-C/d)]$, of unoccupation of each of its
neighbours (say neighbour $i$) in the $Md$ sites problem.
Assuming a replica symmetric (RS) structure, the
 order parameter is the probability  $\mathcal{P}(x)$ that a
randomly chosen site $i$ has $x_i=x$, which satisfies 
the self-consistent equation:
\begin{equation}\label{eq:cavitymatching}
\begin{split}
&\mathcal{P}(x)=\sum_{k=0}^\infty
\frac{C^ke^{-C}}{k!}\int_0^{C}\prod_{a=1}^k\frac{d\xi_a}{C}\int\prod_{a=1}^k\prod_{j_a=1}^{d-1}dx_{j_a}\mathcal{P}(x_{j_a})\\
&\delta\left[x+\frac{1}{\beta}\ln\left(e^{-\beta C/d}+\sum_{a=1}^ke^{-\beta(\xi_a-\sum_{j_a=1}^{d-1}x_{j_a})}\right)\right].
\end{split}
\end{equation}
The free energy $f_{\rm
rs}(\beta)$ can be obtained from  $\mathcal{P}(x)$ as:
\begin{equation}
\begin{split}
f_{\rm rs}(\beta)=&-\frac{d}{\beta}\Big \langle\ln\left(e^{-\beta C/d}+\sum_{a=1}^ke^{-\beta(\xi_a-\sum_{j_a=1}^{d-1}x_{j_a})}\right)\Big\rangle\\
&+\frac{(d-1)C}{\beta}\big\langle \ln\left(1+e^{-\beta(\xi-\sum_{j=1}^dx_j)}\right)\big\rangle
\end{split}
\end{equation}
where $\langle.\rangle$ stands for the averages of the cavity fields $x$ with
the distribution $\mathcal{P}$, of the connectivities $k$ with the Poissonian
distribution of mean $C$ and of the truncated costs $\xi$ with the uniform
distribution in $[0,C]$, as in (\ref{eq:cavitymatching}). In the zero temperature limit, $\beta\to\infty$, we
obtain formulae for the ground state energy that directly generalize the ones
of the BMP case \cite{KrauthMezard89,Aldous01}.

\begin{figure}
\includegraphics[width=8cm, height=5cm,angle=0]{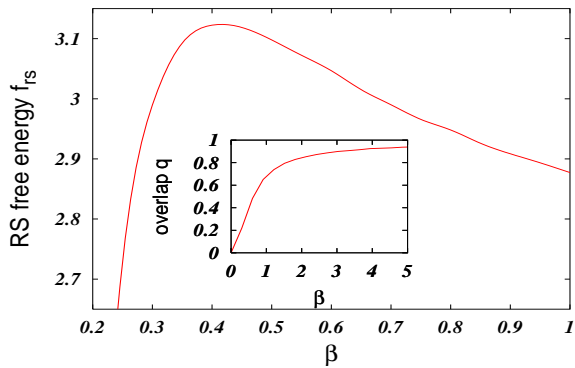}
\caption{Free energy density as a function of inverse temperature $\beta$
in the 3-index matching problem from a population dynamics 
resolution~\cite{MezardParisi01} of the RS 
cavity equations~(\ref{eq:cavitymatching}) with a large enough
value of $C$. (Here $C=60$.).
Note that the entropy $s=\beta^2\partial f/\partial\beta$
is negative for $\beta\geq\beta_s= 0.412\pm 0.001$.
Inset: overlap $q$ between equilibrium configurations as a function of $\beta$; in the glassy phase, the overlap is given by $q(\beta_s)=0.321\pm 0.002$. 
\protect\label{fig:freeenergy}} 
\end{figure}

However, while correctly describing the $d=2$ problem, these RS equations are
inconsistent when $d>2$. We shall discuss specifically the $d=3$ case. First,
the entropy becomes negative for $\beta>\beta_s= 0.412\pm 0.001$ , as shown on
Fig.~\ref{fig:freeenergy}. Second, we have found the RS solution to be
unstable for $\beta>\beta_i$, with $\beta_i\simeq 0.6$ \cite{papierlong}. These two facts show that a discontinuous phase
transition takes place at some inverse temperature $\beta_c\leq\beta_s$. Such
transitions are also present in other NP-hard combinatorial optimization
problems like $K$-SAT, and are usually overcome by passing to a one-step
replica symmetry broken (1RSB) formalism \cite{MezardParisi87b}. Here however,
the direct application of the 1RSB cavity method at zero temperature
\cite{MezardParisi02} turns out to be inadapted.

The originality of the MIMP comes from the peculiar nature of the low
temperature phase. This phase is dominated by isolated configurations, instead
of clusters of configurations that generally arise in 1RSB
systems~\cite{MezardParisi87b}: the 1RSB clusters have no internal entropy
here, a situation which is also found in some other disordered systems, the REM
(random energy model~\cite{Derrida80}), the directed polymer on 
disordered tree~\cite{DerridaSpohn88} and the binary 
perceptron~\cite{KrauthMezard89b}. Upon cooling, these systems freeze 
 when 
reaching the temperature $1/\beta_s$ where the entropy becomes zero.
As a result, the thermodynamical properties can be derived from the
knowledge of the RS solution only~\cite{papierlong}. The free energy is given
by:
\begin{equation} 
f(\beta)= 
\begin{cases} 
f_{\rm rs}(\beta) & {\rm if\ }\beta\leq\beta_s, \\
f_{\rm rs}(\beta_s) & {\rm if\ }\beta\geq\beta_s.
\end{cases} 
\label{free_en}
\end{equation}

Necessary conditions for this {\it frozen} 1RSB Ansatz to hold include the
existence of a finite $\beta_s$ where the RS entropy becomes negative, the
stability of the RS solution up to (at least) $\beta_s$ and the absence of any
discontinuous 1RSB transition before $\beta_s$ (as we have checked from a
finite $\beta$ 1RSB population dynamics investigation).

On top of these properties, a crucial necessary condition for the frozen
1RSB Ansatz to hold is that the system must be subject to a restricted class
of constraints, that we call {\it hard constraints}~\cite{papierlong}. For
matching problems, hard constraints reflect the requirement to realize {\it
perfect} matchings and basically mean that the occupancy of a hyperedge is
uniquely determined by that of its neighbors; this is to be contrasted with
the situation in coloring for instance, where the color of a site is not
necessarily uniquely prescribed by the colors of its neighbors. Notice that
the $d=2$ case satisfies all these requirements, except for the fact that
$\beta_s=\infty$.

The prediction (\ref{free_en}) yields a ground state energy density
$\overline{E_0}/M=f_{\rm rs}(\beta_s)=3.126\pm 0.002$ (see
Fig.~\ref{fig:freeenergy}); our B\&B numerical
estimate is compatible with this value
considering the systematic effects arising from the small $M$
used there.
When $d=4$, we find similarly a ground state energy density
$\overline{E_0}/M=7.703\pm 0.002$ (with $\beta_s=0.135\pm 0.002$);
here again the B\&B estimate we obtained is close to this value.
 
\paragraph*{Overlaps ---}

The cavity method gives  access to the typical overlap $q$ between 
equilibrium configurations,  defined as
\begin{equation}
q=\frac{1}{M}\sum_{i_1,\dots,i_d}\overline{\langle n_{i_1,\dots,i_d}\rangle^2}
\end{equation}
with $\langle.\rangle$ and the overline denoting respectively the 
thermal and the disorder averages. This overlap 
can be expressed with the cavity method in terms of the 
order parameter $\mathcal{P}(x)$.
For $d=3$, we find $q(\beta_s)=0.321\pm 0.002$. 
Because of the special nature of the frozen 1RSB phase
at $\beta>\beta_s$, we expect that, if we take at random
two configurations among the $r$ lowest energy configurations,
their overlap will be equal to $q(\beta_s)$ with probability one
(for any fixed $r$, in the large $M$ limit).

\begin{figure}
\includegraphics[width=6cm, height=8.5cm,angle=270]{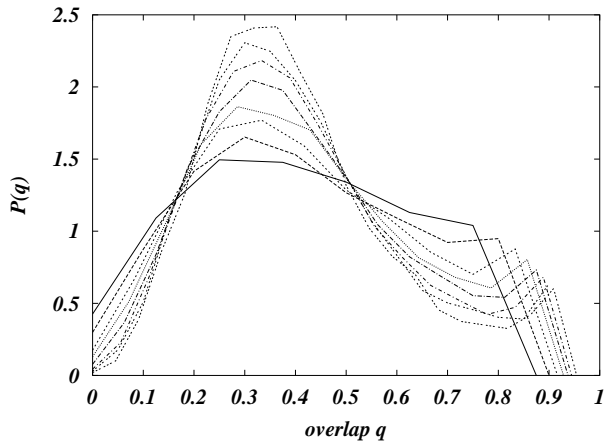}
\caption{Distribution of overlaps between the ground state
and the first excited state for $d=3$ MIMP for  $M=8$ to $M=22$ (from
bottom to top at $q=.32$).
\protect\label{fig:overlaps}}
\end{figure}

In order to test this prediction, we have generalized the B\&B
 method to get numerically the overlap between the ground state
and the first excited state.  Fig.~\ref{fig:overlaps}
 shows  the disorder averaged distribution of
the overlap. The data is consistent
with a distribution becoming peaked at large $M$ at an overlap
around $0.32$, as theoretically predicted
from the cavity approach. Note also that the overlaps 
at higher values seem to decay 
to zero: this is exactly the prediction of the absence of
configurational entropy, i.e., a consequence of the 
freezing scenario which we argued arises in this system.

Another numerical check  of the validity of the scenario
comes from the measurement of the
density ${\cal N}(E)$ of configurations as a function of energy. 
When $(E-E_0)/M$ is small,
we find that $\overline{\ln {\cal N}(E)} \simeq \ln\overline{ {\cal N}(E)} \simeq \gamma (E-E_0)$, with $\gamma(d=3)  \simeq 0.405$, and  
$\gamma(d=4) \simeq 0.14$. These values of $\gamma$ agree with the
inverse freezing temperature $\beta_s$ found in the cavity method.

\paragraph*{Discussion ---}

We have investigated the thermodynamics of the $d$-index matching problem. For
$d \ge 3$ it differs from the (2-index) matching in being NP-hard and having a
low temperature glassy phase. Physically, in the latter case it is much more
difficult to find a second low energy configuration in the neighborhood of a
first one. It would be interesting to study this effect further along the
lines of \cite{HoudayerMartin98,AldousPercus03}. The glassy phase is of a
special type, distinct from the one found in other recently solved NP-complete
decision problems, because it has vanishing internal entropy. In this respect,
the MIMP behaves as a REM~\cite{Derrida80}, freezing into a few
configurations.

We have derived the full phase diagram; we conjecture these results to be
exact, and the numerical checks which we have performed on relatively small
systems, through an efficient B\&B algorithm, are consistent with the
predictions. It will be extremely interesting to generalize to this problem the
rigorous mathematical methods developed for the BMP.

This work was supported in part by the European Community's
Human Potential Programme under contracts 
HPRN-CT-2002-00307 (DYGLAGEMEM) and
HPRN-CT-2002-00319 (STIPCO) as well as by the Community's
EVERGROW Integrated Project.

\bibliographystyle{apsrev}

\bibliography{references,matchings,glasses}

\begin{thebibliography}{21}
\bibitem[{foo()}]{footnote1}
\bibinfo{note}{We must then allow for unoccupied sites: we introduce a chemical
  potential $\mu=C/d$, conjugate to the number of occupied sites, which is sent
  to $\infty$ with $C$ at the end of the computation.}
\end{thebibliography}

\end{document}